\documentclass[twoside,twocolumn,9pt]{article}
\usepackage{extsizes}
\usepackage[super,sort&compress,comma]{natbib} 
\usepackage[version=3]{mhchem}
\usepackage[left=1.5cm, right=1.5cm, top=1.785cm, bottom=2.0cm]{geometry}
\usepackage{balance}
\usepackage{mathptmx}
\usepackage{sectsty}
\usepackage{graphicx} 
\usepackage{lastpage}
\usepackage[format=plain,justification=justified,singlelinecheck=false,font={stretch=1.125,small,sf},labelfont=bf,labelsep=space]{caption}
\usepackage{float}
\usepackage{fancyhdr}
\usepackage{fnpos}
\usepackage[english]{babel}
\addto{\captionsenglish}{%
  
}
\usepackage{array}
\usepackage{droidsans}
\usepackage{charter}
\usepackage[T1]{fontenc}
\usepackage[usenames,dvipsnames]{xcolor}
\usepackage{setspace}
\usepackage[compact]{titlesec}
\usepackage{hyperref}

\usepackage{epstopdf}

\definecolor{cream}{RGB}{222,217,201}
\hypersetup{colorlinks=true,linkcolor=black,filecolor=black,urlcolor=black,citecolor=black}
\begin{document}

\pagestyle{fancy}
\thispagestyle{plain}
\fancypagestyle{plain}{
\renewcommand{\headrulewidth}{0pt}
}

\makeFNbottom
\makeatletter
\renewcommand\LARGE{\@setfontsize\LARGE{15pt}{17}}
\renewcommand\Large{\@setfontsize\Large{12pt}{14}}
\renewcommand\large{\@setfontsize\large{10pt}{12}}
\renewcommand\footnotesize{\@setfontsize\footnotesize{7pt}{10}}
\makeatother

\renewcommand{\thefootnote}{\fnsymbol{footnote}}
\renewcommand\footnoterule{\vspace*{1pt}%
\color{cream}\hrule width 3.5in height 0.4pt \color{black}\vspace*{5pt}} 
\setcounter{secnumdepth}{5}

\makeatletter 
\renewcommand\@biblabel[1]{#1}            
\renewcommand\@makefntext[1]%
{\noindent\makebox[0pt][r]{\@thefnmark\,}#1}
\makeatother 
\renewcommand{\figurename}{\small{Fig.}~}
\sectionfont{\sffamily\Large}
\subsectionfont{\normalsize}
\subsubsectionfont{\bf}
\setstretch{1.125} 
\setlength{\skip\footins}{0.8cm}
\setlength{\footnotesep}{0.25cm}
\setlength{\jot}{10pt}
\titlespacing*{\section}{0pt}{4pt}{4pt}
\titlespacing*{\subsection}{0pt}{15pt}{1pt}

\fancyfoot{}
\fancyfoot[LO,RE]{\vspace{-7.1pt}\includegraphics[height=9pt]{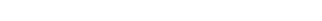}}
\fancyfoot[CO]{\vspace{-7.1pt}\hspace{13.2cm}\includegraphics{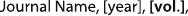}}
\fancyfoot[CE]{\vspace{-7.2pt}\hspace{-14.2cm}\includegraphics{head_foot/RF}}
\fancyfoot[RO]{\footnotesize{\sffamily{1--\pageref{LastPage} ~\textbar  \hspace{2pt}\thepage}}}
\fancyfoot[LE]{\footnotesize{\sffamily{\thepage~\textbar\hspace{3.45cm} 1--\pageref{LastPage}}}}
\fancyhead{}
\renewcommand{\headrulewidth}{0pt} 
\renewcommand{\footrulewidth}{0pt}
\setlength{\arrayrulewidth}{1pt}
\setlength{\columnsep}{6.5mm}
\setlength\bibsep{1pt}

\makeatletter 
\newlength{\figrulesep} 
\setlength{\figrulesep}{0.5\textfloatsep} 

\newcommand{\topfigrule}{\vspace*{-1pt}%
\noindent{\color{cream}\rule[-\figrulesep]{\columnwidth}{1.5pt}} }

\newcommand{\botfigrule}{\vspace*{-2pt}%
\noindent{\color{cream}\rule[\figrulesep]{\columnwidth}{1.5pt}} }

\newcommand{\dblfigrule}{\vspace*{-1pt}%
\noindent{\color{cream}\rule[-\figrulesep]{\textwidth}{1.5pt}} }

\makeatother

\twocolumn[
  \begin{@twocolumnfalse}
{\includegraphics[height=30pt]{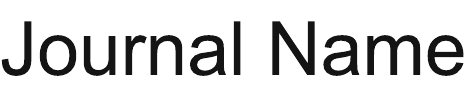}\hfill\raisebox{0pt}[0pt][0pt]{\includegraphics[height=55pt]{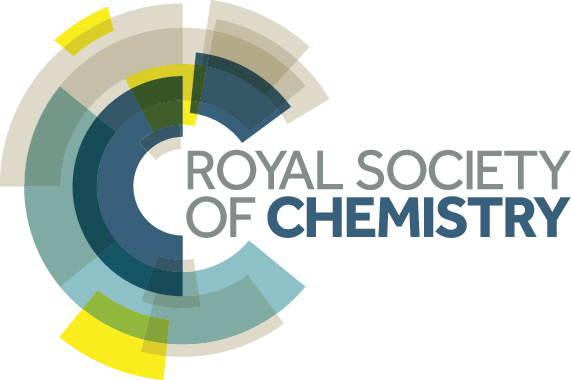}}\\[1ex]
\includegraphics[width=18.5cm]{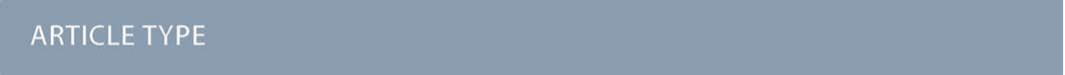}}\par
\vspace{1em}
\sffamily
\begin{tabular}{m{4.5cm} p{13.5cm} }

\includegraphics{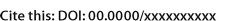} & \noindent\LARGE{\textbf{Expanded-plane bilayer thermal concentrator for improving thermoelectric conversion efficiency}} \\
\vspace{0.3cm} & \vspace{0.3cm} \\

 & \noindent\large{Haohan Tan,\textit{$^{\ddag a}$} Yuqian Zhao,\textit{$^{\ddag a}$} Xinchen Zhou,$^{\ast}$\textit{$^{ab}$} and Jiping Huang$^{\ast}$\textit{$^{a}$}} \\

\includegraphics{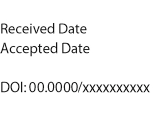} & \noindent\normalsize{Thermoelectric devices are pivotal in the energy sector, with enhancing their conversion efficiency being a longstanding focal point. While progress has been made, overcoming the inherent low efficiency and heat management issues remains challenging. The advent of thermal metamaterials, particularly thermal concentrators, holds promise for improved thermoelectric efficiency. The concentrator has the potential to amplify the temperature gradient within the working region without altering the temperature gradient of the background, thereby enhancing thermoelectric conversion efficiency through this concentrating effect. Nevertheless, the efficacy of this effect is contingent upon the structural parameters of the concentrator. Systematically investigating the impact of metamaterials on thermoelectric conversion efficiency, particularly in terms of quantifying the enhancement, presents a significant challenge. Additionally, the intrinsic thermal conductivity of the material imposes constraints on the applicability of the concentrator in this regard. In this context, drawing inspiration from the recently proposed passive ultra-conductive heat transport scheme, we have devised expanded-plane bilayer thermal concentrators.  We substantiate the prospective performance of our design through analytical demonstration, further validated through finite-element simulations and experiments. Notably, through direct calculation, we illustrate an efficiency improvement of about 38\% when utilizing the expanded-plane concentrator comparing with not using expanded-plane structure. The expanded-plane geometrical configuration of the outer layer can also attain large-scale value. These findings not only present a novel avenue for the functional transformation of thermal metamaterials but also hold significant implications for the field of thermoelectrics.} \\

\end{tabular}

 \end{@twocolumnfalse} \vspace{0.6cm}

  ]

\renewcommand*\rmdefault{bch}\normalfont\upshape
\rmfamily
\section*{}
\vspace{-1cm}


\footnotetext{\textit{$^{a}$~Department of Physics, State Key Laboratory of Surface Physics, and Key Laboratory of Micro and Nano Photonic Structures (MOE), Fudan University, Shanghai 200438, China. E-mail: jphuang@fudan.edu.cn}}
\footnotetext{\textit{$^{b}$~School of Mechanical Engineering, Institute of Refrigeration and Cryogenics, Engineering Research Center of Solar Power and Refrigeration (MOE), Shanghai Jiao Tong University, Shanghai 200240, China. E-mail: zhouxinchen@sjtu.edu.cn}}


\footnotetext{\ddag~These authors contributed equally to this work.}


\section*{Introduction}
\begin{figure*}[htp]
  \includegraphics[width=1.0\linewidth]{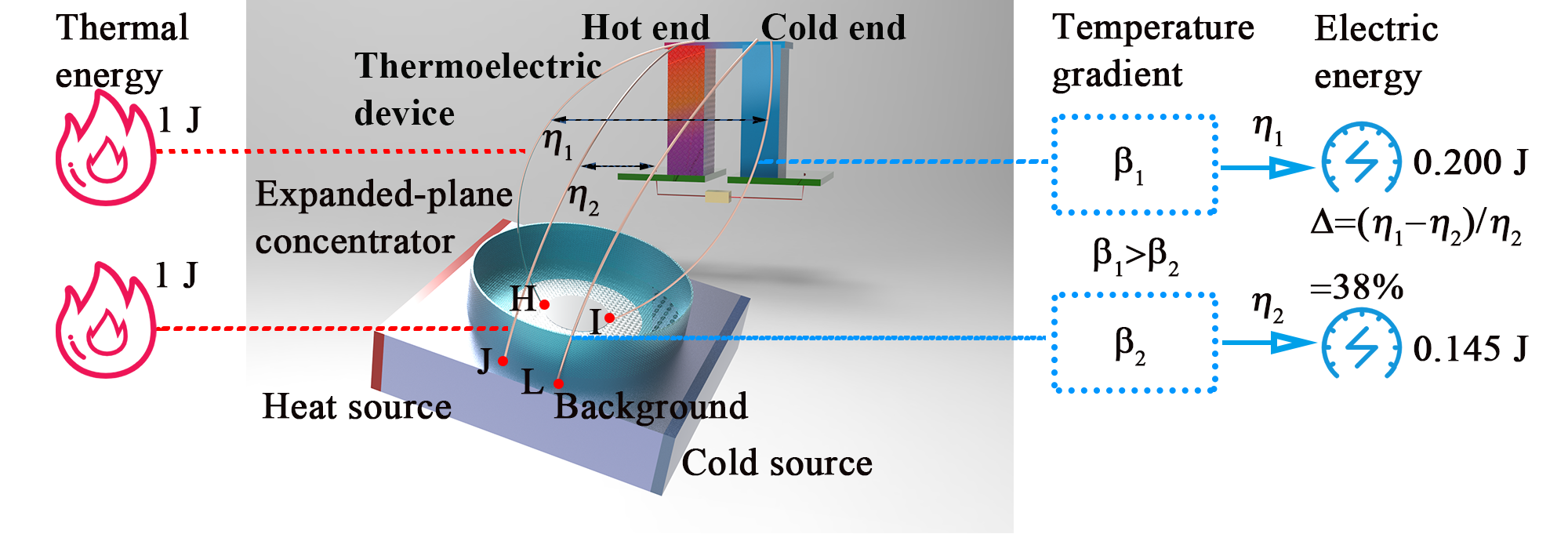}
  \caption{The improvement of thermoelectric conversion efficiency via EP concentrator. The efficiency improvement of the EP concentrator comparing thermal conversion efficiency under two distinct scenarios: $\eta_1$, where the device is in contact with heat and cold sources at the core region's boundary (at points H and I, i.e., utilizing our design), and $\eta_2$, where the sources are situated in boundary between out-layer region and the background region (at points J and L, i.e., withour utilizing our design). $\beta_{1}$ and $\beta_{2}$ are the corresponding thermal gradients.}
  \label{f1}
\end{figure*}

Presently, the issue of energy management has become an increasingly critical challenge. Owing to the limited availability of resources and environmental implications, there is an evident shift towards the utilization of various forms of renewable energy, encompassing solar, wind, and thermal energy, among others. In the realm of thermal energy, the primary focus lies in the efficient conversion to electrical power. In this context, leveraging the Seebeck effect, researchers have developed thermoelectric devices that are capable of directly transforming thermal energy into electrical energy. Additionally, these advancements have led to the development of various applications, notably in waste heat recovery, as exemplified in Li et al.'s research~\cite{li2023realizing}, and in ocean thermal energy conversion, highlighted in Vega et al.'s study~\cite{vega2012ocean}. For thermoelectric devices, thermoelectric conversion efficiency (TCE) is fundamentally linked to the thermoelectric figure of merit, denoted as $zT$. In pursuit of enhancing $zT$, scholars have dedicated efforts to various approaches such as bandgap tuning, chemical potential engineering, and entropy engineering~\cite{ghawri2022breakdown,chen2021leveraging}. Notably, Rafal et al. have reported high thermoelectric conversion efficiency of $10.8\%$ in $\rm Bi_2Te_3$-based stepwise structures through bandgap and chemical potential modifications~\cite{knura2024achieving}. Similarly, Abid et al. have achieved significant enhancements in the thermoelectric performance of p-type $\rm Bi_2Te_3$ materials via entropy engineering~\cite{ahmad2024largely} and realized the TCE of $8\%$. Li et al. also achieved an excellent conversion efficiency of $14.5\%$ in the $\rm Mg_3Sb_2$/GeTe-based thermoelectric module~\cite{li2023realizing}, and the scope of research has expanded to include nanostructures~\cite{novak20192d}. Theoretical work has progressed to encompass more complex scenarios, such as Gao et al.'s consideration of TCE under a constant heat flux as opposed to a constant temperature differential~\cite{min2022new}. Despite significant advancements and improvements in TCE since the early days, the issues of inherently low TCE and collection efficiency remain unresolved.

Metamaterials have emerged as a groundbreaking field over the past decades, with thermal metamaterials constituting a significant segment~\cite{zhang2023diffusion,yangmass,xu2023transformation,huang2020theoretical,xu2021geometric,xu2020transformation1,yang2017full,dai2018transient,xu2018thermal1}. Since Fan et al. introduced the concept of thermal metamaterials in 2008~\cite{fan2008shaped}, the field has seen substantial progress~\cite{li2021transforming,xu2022thermal,xu2022diffusive,jin2023tunable,xu2023giant,zhou2023adaptive,xi2023ultrahigh,xu2020transformation,shen2016thermal1}. These innovative materials have laid the groundwork for devices featuring new functionalities, such as thermal cloaks~\cite{yang2022transformation,zhuang2022nonlinear,wegener2013metamaterials,nguyen2015active,yue2021thermal,yang2021optimization,han2013homogeneous,shen2016thermal}, concentrators~\cite{moccia2014independent,chen2015experimental,xu2019converging,xu2018thermal,ji2021designing,xu2019bilayer,fujii2020cloaking,chen2015materials,sun2022design,guenneau2012transformation}, sensor~\cite{jin2020making}, illusion~\cite{zhu2015converting,yang2019thermal}, transparency~\cite{xu2019thermal}, and rotators~\cite{yang2020experimental,guenneau2013anisotropic}, evolving structurally from single-layer to multilayer configurations~\cite{xu2014ultrathin,xu2019bilayer}. Thermal concentrators, which can increase temperature gradient of core region while keeping temperature gradient of background undistorted, hold promise for enhancing energy management and TCE. Nonetheless, the pursuit of efficiency improvements via thermal concentrators remains insufficiently explored. Investigations into the temperature concentration effect through concentrators have demonstrated near $100\%$ efficiency; however, these studies have not addressed TCE enhancements~\cite{han2013theoretical}. Li et al. have discussed voltage improvements using thermal metamaterials~\cite{li2022energy}, yet their findings are predominantly simulation and experiment-driven, with a theoretical gap remaining. A comprehensive analysis of the impact of thermal metamaterials on TCE remains to be conducted, particularly with respect to quantifying the relationship between TCE and the parameters of the thermal concentrator. This relationship is influenced by variables such as the radius and thermal conductivity of the core region. Additionally, temperature is an important factor that warrants consideration in certain contexts. Furthermore, the reliance on natural materials with inherent thermal conductivity constraints hinders achieving desired performance levels and limits the broader application potential. Although the thermal conductivity of diamond can attain about $2000$ W/(m$\cdot$K), the thermal conductivity of common materials can only reach about $400$ W/(m$\cdot$K)~\cite{guo2022passive}. However, various strategies have been proposed to surpass these natural limitations, including convection and phase change~\cite{li2019mechanism,sigurdson2013large}. A recent passive approach utilizing an expanded plane orthogonal to the two-dimensional plane has been identified, enabling the achievement of an ultrahigh effective thermal conductivity beyond natural material capabilities~\cite{guo2022passive}. Inspired by this approach, a bilayer thermal cloak without interfacial thermal resistance has been conceptualized to realize high thermal conductivity~\cite{han2023itr}. Yet, engineering a multilayer thermal concentrator, such as a bilayer configuration, remains a complex endeavor. Contrary to the cloaking effect, the concentrating effect depends significantly on the properties of the core region.

In this paper, we explore the foundational theory of a bilayer thermal concentrator employing the expanded-plane (EP) structure, deliberately omitting the impact of both the expanded plane and substrate thickness. Although our theoretical framework is elementary, we validate the superior performance of the proposed structure via simulations, demonstrating the concentrating effect with quantitative results. Additionally, by analyzing temperatures at two strategic points on the boundary of the core region and their counterparts in the background region, i.e., using and not using our design, we compute the TCE in these varying contexts. The results indicate a significant efficiency enhancement of approximately $38\%$ when utilizing EP structure, marking a notable advancement in the field of thermoelectrics. These insights not only pave the way for further investigations into multifunctional EP structures but also promise to significantly influence developments in heat management.
\section*{Theory of EP concentrator}
The structure of the EP bilayer concentrator is illustrated in Fig.~\ref{moni}. In this representation, $R_1$, $R_2$, and $R_3$ denote the core radius, inner layer radius, and outer layer radius, respectively. The dimensions of the substrate are characterized by its length and width, denoted as $L$, while the thickness of the substrate is represented by $v$. The expanded plane is characterized by its thickness, denoted as $d$, and its height, represented by $h_3$. The thermal conductivity of the core region, inner layer, outer layer, background region, and the expanded plane is symbolized by $\kappa_1$, $\kappa_2$, $\kappa_3$, $\kappa_4$, and $\kappa_5$, respectively. The corresponding temperature fields for each region are denoted by $T_1$, $T_2$, $T_3$, $T_4$, and $T_5$. We assume that the thickness $v$ is sufficiently small to disregard the temperature gradient of the substrate along the vertical direction. In other words, the substrate can be treated as a pseudo-2D conduction system. According to the isotropic theory of bilayer thermal concentrators, the temperature field of each region can be determined by directly solving the static heat transfer equation, i.e., the Laplace equation.
The temperature distributions in the EP bilayer concentrator are expressed by the following equations:
\begin{subequations}
  \begin{equation}
    T_1 = A r \cos\theta \ (r<R_1),
  \end{equation}
  \begin{equation}
    T_2 = B r \cos\theta + \frac{C \cos\theta}{r} \ (R_1<r<R_2),
  \end{equation}
  \begin{equation}
    T_3 = D r \cos\theta + \frac{E \cos\theta}{r} \ (R_2<r<R_3),
  \end{equation}
  \begin{equation}
    T_4 = F r \cos\theta \ (r>R_3),
  \end{equation}
\end{subequations}
where $A$, $B$, $C$, $D$, $E$, and $F$ are constants dependent on the boundary conditions. Here, $r$ represents the distance between the field point and the origin, and $\theta$ is the polar angle (refer to Fig.~\ref{moni}(a3)). The assumed adiabatic boundaries of the expanded plane exclude the bottom boundary. Due to the continuity of temperature at the boundary between the expanded plane and the substrate, the temperature distribution on the expanded plane is given by:

\begin{equation}
  T_5 = \frac{F R_3 \cos\theta}{\cosh\left(\frac{h_3}{R_3}\right)}\cosh\left(\frac{h_3-z}{R_3}\right).
\end{equation}
Then, further considering other boundary conditions of temperature consistency and flux continuous between two adjacent regions, we can get the following system of equations:
\begin{eqnarray}
  FR_3=DR_3+\frac{E}{R_3},\\
  \kappa_4F=\kappa_5F\tanh\left(\frac{h_3}{R_3}\right)+\kappa_3\left(D-\frac{E}{{R_3}^2}\right),\\
  DR_2+\frac{E}{R_2}=BR_2+\frac{C}{R_2},\\
  \kappa_3\left(D-\frac{E}{{R_2}^2}\right)=\kappa_2\left(B-\frac{C}{{r_2}^2}\right),\\
  BR_1+\frac{C}{R_1}=AR_1,\\
  \kappa_2\left(BR_1+\frac{C}{R_1}\right)=\kappa_1R_1.
\end{eqnarray}
Since the necessary condition for successfully achieving the bilayer thermal concentrator is that the above system of equations has non-zero solutions, we let the determinant of coefficients of the above system of equations be zero. Then, we can know the relation among the parameters.

\section*{Finite-element simulations of EP concentrator}
\begin{figure*}[htp]
  \includegraphics[width=1.0\linewidth]{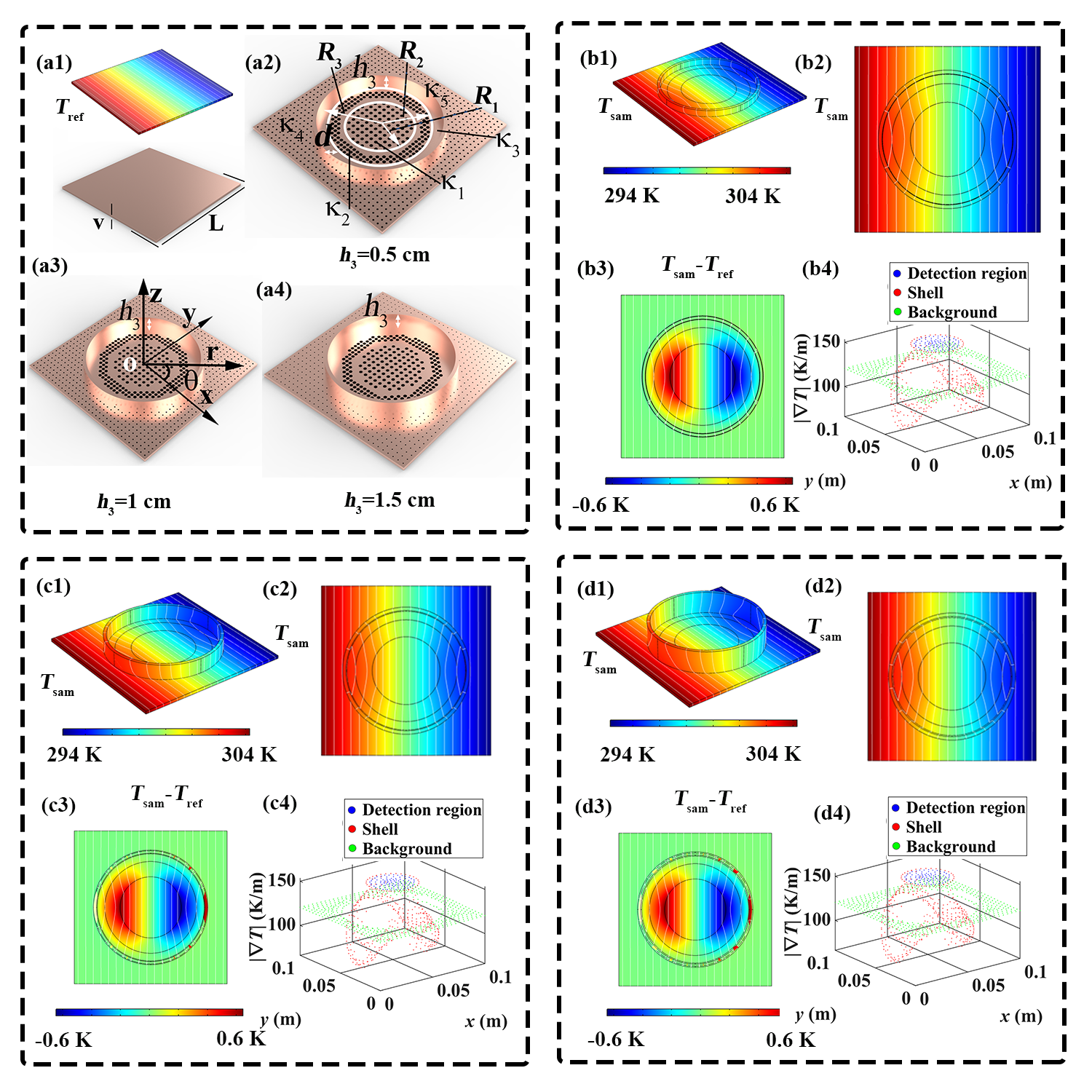}
  \caption{
    The structures of EP concentrators and reference and its simulation results are depicted.
    (a1) The simulation employs the following parameters: $L=10$ cm, $R_1=2$ cm, $R_2=3$ cm, $R_3=3.5$ cm, $v=0.2$ cm, $d=0.2$ cm, $\kappa_1=130$ W/(m$\cdot$K), $\kappa_2=281.5$ W/(m$\cdot$K), $\kappa_3=400$ W/(m$\cdot$K), and $\kappa_5=400$ W/(m$\cdot$K). The thermal conductivities of the background region are as follows:
    (a2) $\kappa_4=301.3$ W/(m$\cdot$K);
    (a3) $\kappa_4=355.8$ W/(m$\cdot$K);
    (a4) $\kappa_4=406.2$ W/(m$\cdot$K).
    The corresponding height of expanded plane is $h_3=0.5$ cm, $h_3=1$ cm and $h_3=1.5$ cm, respectively.
    (b1)-(b4) The simulation presents the temperature distribution, the temperature difference between structure in (a2) and the reference, and the temperature gradient for structure in (a2).
    (c1)-(c4) The corresponding simulation results are shown for structure in (a3).
    (d1)-(d4) The corresponding simulation results are shown for structure in (a4).}
  \label{moni}
\end{figure*}
To validate the aforementioned theory, we employed COMSOL Multiphysics for finite-element simulations~\cite{xu2020active}. Various schemes can be utilized to realize the concentrator based on the relations among the parameters (see chosen parameters in Method).

The simulation results are presented in Fig.\ref{moni}(b1)-(b2). The proposed concentrator maintains undisturbed background isothermal lines while concentrating the heat flux in the core region. These outcomes affirm the excellent performance of the proposed structure in achieving a concentrating effect. To quantitatively assess the concentrator's performance, we display the temperature difference between the EP structure and reference in Fig.~\ref{moni}(b3). Additionally, the temperature gradient of EP structures is depicted in Fig.~\ref{moni}(b4), providing further confirmation of the concentrator's effectiveness.

To further evaluate performance, we adjusted the height of the expanded plane and the thermal conductivity of the background region. The simulation results, shown in Fig.~\ref{moni}(c1)-(c4),(d1)-(d4), illustrate the structure's consistent performance across different expanded plane heights.

\section*{Experimental validation of EP concentrator}
\begin{figure*}[htp]
  \includegraphics[width=1.0\linewidth]{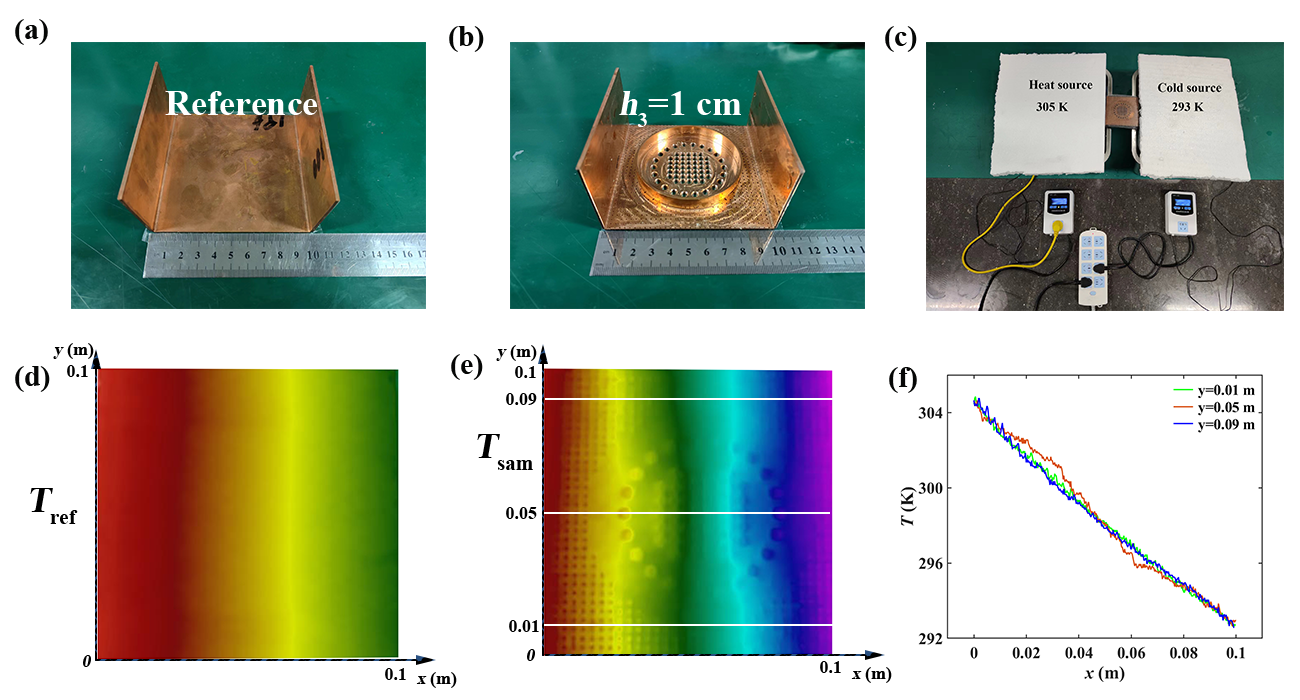}
  \caption{The experimental setup and results. (a) The reference without expanded plane. (b) The sample with expanded plane. (c) The experimental setup.
    (d) The experimental results of reference. (e) The experimental results of EP sample. (f) The temperature distribution of three lines on the bottom surface of EP sample, i.e., $y=0.01~\rm m$, $y=0.05~\rm m$, and $y=0.09~\rm m$. The average temperature gradients of core region ($\beta_{1}$) and background ($\beta_{2}$) are $155~\rm K/m$ and $115~\rm K/m$, respectively.}
  \label{shiyan}
\end{figure*}

To experimentally validate the performance of the EP bilayer concentrator, we fabricated both a reference and an EP sample using copper, as illustrated in Fig.\ref{shiyan}(a)-(b). The parameters selected for the EP sample were in alignment with those depicted in Fig.\ref{moni}(b1) (see specific parameters in Method).

The experimental setup is depicted in Fig.\ref{shiyan}(c). On the left side, the sample was submerged in a hot water tank at $305$ K, temperature-maintained by a heating bar. Conversely, the right side was immersed in cold water at $293$ K, cooled using ice packs. After achieving a stable temperature field within the sample, we recorded the surface temperature distribution using an infrared camera from above. This procedure was similarly conducted for the reference sample, with results presented in Fig.\ref{shiyan}(d)-(e). To quantitatively assess the concentrating effect, we examined the temperature distribution along three specific lines on the sample's bottom surface, namely at $y=0.01~\rm m$, $y=0.05~\rm m$, and $y=0.09~\rm m$, as shown in Fig.~\ref{shiyan}(f). The observations revealed that while the background temperature distribution remained unaltered, the isothermal line was notably concentrated towards the core region. The average temperature gradients of core region ($\beta_{1}$) and background ($\beta_{2}$) are $155~\rm K/m$ and $115~\rm K/m$, respectively. These findings robustly demonstrate the outstanding efficacy of the EP bilayer concentrator.

\begin{table*}
  \small
  \caption{The value of $S$, $\rho$, and $\kappa$ of p-PbTe under different temperature}
  \label{tbl:example1}
  \begin{tabular*}{\textwidth}{@{\extracolsep{\fill}}lllll}
    \hline
    Temperature $T$ (K)&
    Seebeck coefficient $S$ ($10^{-4}\rm V~K^{-1}$)&
    Resistivity $\rho$ ($10^{-5}\Omega$~m)&
    Thermal conductivity $\kappa$ ($\rm W/(m\cdot K)$)&
    $zT$\\
    \hline
    300 & 1.06 & 0.71 & 2.52 & 0.19 \\
    400 & 1.54 & 1.01 & 1.79 & 0.52 \\
    500 & 2.05 & 1.59 & 1.59 & 0.95 \\
    \hline
  \end{tabular*}
\end{table*}

\section*{EP concentrator assisted improvement of thermoelectric conversion efficiency}

Thermoelectric devices, capable of converting thermal energy into electrical energy, have garnered significant attention since their inception. According to thermoelectric theory, the efficiency of such devices can be expressed as:
\begin{eqnarray}
  \eta=\frac{T_h-T_c}{T_h}\frac{\sqrt{zT+1}-1}{\sqrt{zT+1}+T_c/T_h},
  \label{themo}
\end{eqnarray}
where $T_h$, $T_c$, and $T$ denote the temperatures of the heat source, cold source, and the average temperature, respectively. The parameter $z$ is related to the Seebeck coefficient $S$, electrical resistivity $\rho$, and thermal conductivity $\kappa$ through the equation $z=\frac{S^2}{\rho\kappa}$.
\begin{figure*}[htp]
  \includegraphics[width=1.0\linewidth]{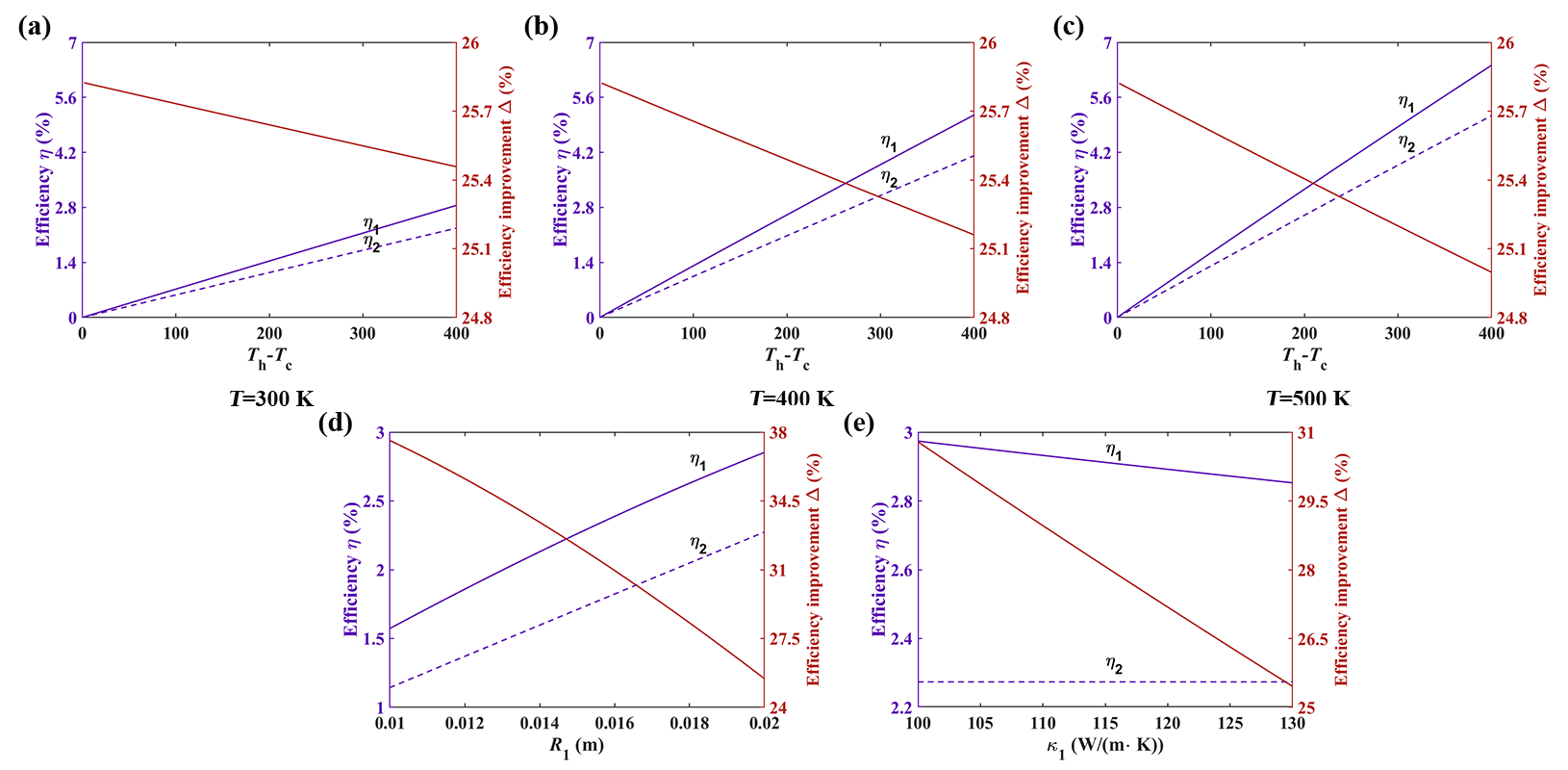}
  \caption{The quantitative results of improvement for thermoelectric conversion efficiency via EP concentrator. The variation of efficiency $\eta_1$ using EP concentrator, i.e., assuming the heat and cold source locating at boundary of core region, the efficiency $\eta_2$ without using EP concentrator, i.e., assuming the heat and cold source locating at background region, and the percentage change of efficiency comparing $\eta_2$ and $\eta_1$ with: (a) temperature difference under average temperature 300 K; (b) temperature difference under average temperature 400 K;  (c) temperature difference under average temperature 500 K; (d) radius $R_1$ of core region; (e) thermal conductivity $\kappa_1$ of core region. }
  \label{fig4}
\end{figure*}
To elucidate the efficiency enhancement afforded by the concentrator, we computed the efficiency under two distinct scenarios: $\eta_1$, where the device is in contact with heat and cold sources at the core region's boundary (at points H and I), and $\eta_2$, where the sources are situated in background region (at points J and L). We considered three different average temperatures—300 K, 400 K, and 500 K—while adjusting $T_h$ and $T_c$. The parameters for these scenarios and the corresponding $zT$ values are detailed in Table~\ref{tbl:example1}. Utilizing Eq.(\ref{themo}), we derived the efficiencies $\eta_1$ and $\eta_2$, as depicted in Fig.\ref{fig4}. Notably, both $\eta_1$ and $\eta_2$ demonstrate an increase with the growing temperature differential $T_h-T_c$. When specifically examining the percentage change in efficiency, our analysis reveals that the maximum efficiency increase can reach approximately $25.8\%$, signifying a substantial impact.

Additionally, we explored the influence of $R_1$ on TCE. Both $\eta_1$ and $\eta_2$ exhibited an uptrend, while $\Delta$ decreased as $R_1$ increased, with $\Delta$ achieving a peak value of nearly $38\%$. The variations of $\eta_1$, $\eta_2$, and $\Delta$ in relation to $\kappa_1$ are also presented in Fig.~\ref{fig4}(e). We observed that while $\eta_2$ remained constant, both $\eta_1$ and $\Delta$ decreased with an increase in $\kappa_1$, reaching a maximum $\Delta$ of about $31\%$. These findings significantly highlight the prospective applications of thermal concentrators in thermoelectric technology.
\section*{DISCUSSION AND CONCLUSION}
We have introduced a novel structure to achieve a bilayer thermal concentrator, utilizing an expanded plane extending perpendicularly to the two-dimensional plane. Both simulations and experiments confirm the exceptional performance of this proposed structure. In comparison with conventional bilayer thermal concentrators, the EP-bilayer-concentrator exhibits distinctive features, with a crucial advantage being the potential to achieve a very high effective thermal conductivity in the outer-layer region ($\kappa_{3,eff}$). For instance, considering the EP structure in Fig.~\ref{moni}(c1), the effective thermal conductivity of the outer-layer region is calculated to be $1071$~W/(m$\cdot$K), according to conventional bilayer concentrator theory\cite{xu2019bilayer} (see calculating process in Method). This property overcomes the limitations imposed by the relatively low thermal conductivity of natural materials, expanding the potential applications of bilayer thermal concentrators. Additionally, by directly calculating the efficiency of a thermoelectric device situated in the core region and background region, we substantiate the efficiency improvement conferred by the thermal concentrator.

However, it is noteworthy that our research overlooks the impact of thermal convection and thermal radiation. Future investigations should consider the influence of these factors. Despite this, the proposed structure exhibits satisfactory performance even when neglecting the effects of $d$ and $v$, attributable to the relatively small size of our sample. For larger structures, especially those with significant $d$ and $v$ values, the concentrating effect may not be as pronounced. Multilayer structures and enhanced heat management with multiple expanded planes can also be explored~\cite{imran2020advanced}. Furthermore, in our efficiency calculations for thermoelectric devices, we assumed constant values for $S$, $\rho$, and $\kappa$. Future studies should consider the influence of temperature on device parameters.

In summary, our study introduces an innovative structure designed for a bilayer thermal concentrator. We established the interrelationships among various parameters using an approximate theory, intentionally overlooking the thickness of both the expanded plane and the substrate. The validity of our theoretical approach is corroborated by finite-element simulations and experimental evidence, which collectively affirm the impressive concentrating capabilities of our proposed structure. In comparison to traditional two-dimensional concentrators, our design demonstrates versatility across diverse environments with varying thermal conductivities. Furthermore, we have conducted a systematic investigation into the efficiency variations of thermoelectric conversion associated with each parameter. The substantial improvement in efficiency observed for thermoelectric devices using our concentrator underscores the significance of our findings, potentially revolutionizing research in thermal metamaterials and heat management. Our research is poised to make profound contributions to both the energy and thermoelectric materials fields. There is an expectation that, by integrating microscopic and macroscopic approaches, thermal conversion efficiency can be further augmented. Consequently, this advancement will enable more efficient conversion of thermal energy into electrical energy, which could play a crucial role in alleviating global energy challenges~\cite{gao2007magnetophoresis,dong2004dielectric,huang2003dielectrophoresis,ye2008non,liu2013statistical,huang2005magneto,qiu2015nonstraight}.

\section*{Methods}
\subsection*{The chosen parameters for EP concentrator}
In our initial simulation, we set the values of $R_1$, $R_2$, $R_3$, $L$, and $h_3$ as follows: $R_1=2$ cm, $R_2=3$ cm, $R_3=3.5$ cm, $L=10$ cm, and $h_3=0.5$ cm. Ensuring the concentrating effect, the thermal conductivity of each region was set as follows: $\kappa_1=130$ W/(m$\cdot$K), $\kappa_2=281.5$ W/(m$\cdot$K), $\kappa_3=400$ W/(m$\cdot$K), $\kappa_4=301.3$ W/(m$\cdot$K), and $\kappa_5=400$ W/(m$\cdot$K). The left boundary was in contact with a heat source at a temperature of $304$ K, and the right boundary with a cold source at $294$ K. Other boundaries were set as adiabatic. We adjusted the height of the expanded plane ($h_3$) by specifically setting it to 1 cm and 1.5 cm. The thermal conductivity of the background region ($\kappa_4$) was adjusted to $355.8$ W/(m$\cdot$K) and $406.2$ W/(m$\cdot$K), respectively.
\subsection*{The chosen parameters and fabrication of EP sample }
For the EP sample, the following parameters were set: $L=10$ cm, $R_1=2$ cm, $R_2=3$ cm, $R_3=3.5$ cm, $h=10$ cm, $\kappa_1=130$ W/(m$\cdot$K), $\kappa_2=281.5$ W/(m$\cdot$K), $\kappa_3=400$ W/(m$\cdot$K), $\kappa_4=355.8$ W/(m$\cdot$K). To achieve the desired thermal conductivity in each region, we employed air holes drilled in each substrate region based on the effective medium theory~\cite{choy2015effective,davis1977effective,tian2021thermal}. This design eliminates interfacial thermal resistance between adjacent regions. Given the thermal conductivity of copper and air as $400$ W/(m$\cdot$K) and $0.03$ W/(m$\cdot$K), respectively, holes were not drilled in the outer-layer region. The area fractions for the core region, inner-layer region, and background region were set to $50\%$, $17\%$, and $5\%$, respectively. With hole diameters in the core region, inner-layer region, and background region at $2\times10^{-3}$ m, $2\times10^{-3}$ m, and $4\times10^{-4}$ m, respectively, the number of holes in each region were $50$, $22$, and $716$.
\subsection*{The calculation of effective thermal conductivity for the outer-layer region}
According to the theory of conventional two-dimensional bilayer concentrator, the relationship between thermal conductivity of outer-layer region $\kappa_3$ and other parameters can be expressed as follows:
\begin{align}
  \kappa_{3,eff} & =-\left[\left(R_1^4R_2^4\kappa_1^2\kappa_2^2 - 2R_1^4R_2^4\kappa_1^2\kappa_2\kappa_4 + R_1^4R_2^4\kappa_1^2\kappa_4^2\nonumber  \right.\right.  \\
           & - 2R_1^4R_2^4\kappa_1\kappa_2^3 + 4R_1^4R_2^4\kappa_1\kappa_2^2\kappa_4 - 2R_1^4R_2^4\kappa_1\kappa_2\kappa_4^2\nonumber \\
           & + R_1^4R_2^4\kappa_2^4 - 2R_1^4R_2^4\kappa_2^3\kappa_4 + R_1^4R_2^4\kappa_2^2\kappa_4^2 + 2R_1^4R_2^2R_3^2\kappa_1^2\kappa_2^2\nonumber                        \\
           & + 12R_1^4R_2^2R_3^2\kappa_1^2\kappa_2\kappa_4 + 2R_1^4R_2^2R_3^2\kappa_1^2\kappa_4^2
           - 4R_1^4R_2^2R_3^2\kappa_1\kappa_2^3\nonumber                            \\
           & - 24R_1^4R_2^2R_3^2\kappa_1\kappa_2^2\kappa_4- 4R_1^4R_2^2R_3^2\kappa_1\kappa_2\kappa_4^2
           + 2R_1^4R_2^2R_3^2\kappa_2^4\nonumber                                         \\
           & 
  + 12R_1^4R_2^2R_3^2\kappa_2^3\kappa_4+ 2R_1^4R_2^2R_3^2\kappa_2^2\kappa_4^2
  + R_1^4R_3^4\kappa_1^2\kappa_2^2\nonumber                                                                                      \\
           &  - 2R_1^4R_3^4\kappa_1^2\kappa_2\kappa_4+ R_1^4R_3^4\kappa_1^2\kappa_4^2- 2R_1^4R_3^4\kappa_1\kappa_2^3\nonumber                                                   \\
           &  + 4R_1^4R_3^4\kappa_1\kappa_2^2\kappa_4- 2R_1^4R_3^4\kappa_1\kappa_2\kappa_4^2+ R_1^4R_3^4\kappa_2^4\nonumber         \\
           & 
  - 2R_1^4R_3^4\kappa_2^3\kappa_4+ R_1^4R_3^4\kappa_2^2\kappa_4^2+ 2R_1^2R_2^6\kappa_1^2\kappa_2^2\nonumber                                                                                            \\
           &  - 2R_1^2R_2^6\kappa_1^2\kappa_4^2 - 2R_1^2R_2^6\kappa_2^4+ 2R_1^2R_2^6\kappa_2^2\kappa_4^2\nonumber            \\
           &  + 4R_1^2R_2^4R_3^2\kappa_1^2\kappa_2^2- 4R_1^2R_2^4R_3^2\kappa_1^2\kappa_4^2 - 4R_1^2R_2^4R_3^2\kappa_2^4\nonumber                 \\
           & 
  + 4R_1^2R_2^4R_3^2\kappa_2^2\kappa_4^2+ 2R_1^2R_2^2R_3^4\kappa_1^2\kappa_2^2 - 2R_1^2R_2^2R_3^4\kappa_1^2\kappa_4^2\nonumber                                                                                     \\
           &  - 2R_1^2R_2^2R_3^4\kappa_2^4+ 2R_1^2R_2^2R_3^4\kappa_2^2\kappa_4^2+ R_2^8\kappa_1^2\kappa_2^2 \nonumber      \\
           & + 2R_2^8\kappa_1^2\kappa_2\kappa_4+ R_2^8\kappa_1^2\kappa_4^2+ 2R_2^8\kappa_1\kappa_2^3+ 4R_2^8\kappa_1\kappa_2^2\kappa_4\nonumber            \\
           & + 2R_2^8\kappa_1\kappa_2\kappa_4^2+ R_2^8\kappa_2^4 + 2R_2^8\kappa_2^3\kappa_4+ R_2^8\kappa_2^2\kappa_4^2+ 2R_2^6R_3^2\kappa_1^2\kappa_2^2\nonumber                                  \\
           & 
  - 12R_2^6R_3^2\kappa_1^2\kappa_2\kappa_4 + 2R_2^6R_3^2\kappa_1^2\kappa_4^2+ 4R_2^6R_3^2\kappa_1\kappa_2^3 - 24R_2^6R_3^2\kappa_1\kappa_2^2\kappa_4\nonumber                                                                                   \\
           & + 4R_2^6R_3^2\kappa_1\kappa_2\kappa_4^2+ 2R_2^6R_3^2\kappa_2^4 - 12R_2^6R_3^2\kappa_2^3\kappa_4+ 2R_2^6R_3^2\kappa_2^2\kappa_4^2\nonumber                 \\
           & + R_2^4R_3^4\kappa_1^2\kappa_2^2 + 2R_2^4R_3^4\kappa_1^2\kappa_2\kappa_4+ R_2^4R_3^4\kappa_1^2\kappa_4^2+ 2R_2^4R_3^4\kappa_1\kappa_2^3\nonumber       \\
           & 
  + 4R_2^4R_3^4\kappa_1\kappa_2^2\kappa_4+ 2R_2^4R_3^4\kappa_1\kappa_2\kappa_4^2+ R_2^4R_3^4\kappa_2^4 + 2R_2^4R_3^4\kappa_2^3\kappa_4\nonumber                                                                                    \\           
           &\left.+ R_2^4R_3^4\kappa_2^2\kappa_4^2\right)^{1/2}- R_2^4\kappa_2^2 + R_1^2R_2^2\kappa_2^2+ R_1^2R_3^2\kappa_2^2 - R_2^2R_3^2\kappa_2^2\nonumber                                 \\
           & - R_2^4\kappa_1\kappa_2+ R_2^4\kappa_1\kappa_4 + R_2^4\kappa_2\kappa_4 - R_1^2R_2^2\kappa_1\kappa_2- R_1^2R_3^2\kappa_1\kappa_2\nonumber                                            \\
           &  - R_1^2R_2^2\kappa_1\kappa_4- R_2^2R_3^2\kappa_1\kappa_2+ R_1^2R_2^2\kappa_2\kappa_4 - R_1^2R_3^2\kappa_1\kappa_4 \nonumber                                    \\  
           & \left.+ R_1^2R_3^2\kappa_2\kappa_4+ R_2^2R_3^2\kappa_1\kappa_4+ R_2^2R_3^2\kappa_2\kappa_4\right]/\left[2\left(R_2^4\kappa_1 + R_2^4\kappa_2\nonumber\right.\right.                          \\
           &\left.\left. - R_1^2R_2^2\kappa_1 + R_1^2R_2^2\kappa_2 + R_1^2R_3^2\kappa_1- R_1^2R_3^2\kappa_2 - R_2^2R_3^2\kappa_1 - R_2^2R_3^2\kappa_2\right)\right].\nonumber                  \\               
\end{align}
Substituting the parameters of EP structure with expanded plane of 1 cm, we can obtain the effective thermal conductivity of outer-layer region: $\kappa_{3,eff}=1071\rm~W/(m\cdot K)$.

\section*{Conflicts of interest}
There are no conflicts to declare.

\section*{Acknowledgements}
We gratefully acknowledge funding from the National Natural Science Foundation of China (Grants No. 12035004 and No. 12320101004) and the Innovation Program of Shanghai Municipal Education Commission (Grant No. 2023ZKZD06).



\balance


\providecommand*{\mcitethebibliography}{\thebibliography}
\csname @ifundefined\endcsname{endmcitethebibliography}
{\let\endmcitethebibliography\endthebibliography}{}

\bibliographystyle{rsc} 

\end{document}